\documentclass[%
reprint,
nofootinbib,
 amsmath,amssymb,
 aps,
]{revtex4-2}

\usepackage{graphicx}
\usepackage{dcolumn}
\usepackage{bm}
\usepackage{color}

\usepackage{subfigure}

\begin{document}

\preprint{APS/123-QED}

\title{Seismic background mitigation with the Lunar Gravitational-wave Antenna}

\author{Han Yan$^{1,2,3}$}
    \email{Corresponding author. \\ yanhanphy@pku.edu.cn, hyan.phy@gmail.com}
\author{Jan Harms$^{3,4}$}

\affiliation{$^{1}$Department of Astronomy, School of Physics, Peking University, 100871 Beijing, China}

\affiliation{$^{2}$Kavli Institute for Astronomy and Astrophysics, Peking University, 100871 Beijing, China}

 \affiliation{$^{3}$Gran Sasso Science Institute, 67100 L'Aquila, Italy}

\affiliation{$^{4}$INFN, Laboratori Nazionali del Gran Sasso, 67100 Assergi, Italy}

\date{\today}

\begin{abstract}
Lunar gravitational-wave (GW) detectors relying on the measurement of the response of the Moon to GWs are susceptible to a seismic background, which might pose a fundamental sensitivity limitation. The Lunar Gravitational-wave Antenna (LGWA) was conceived as an array of accelerometers with the idea that data can be processed to distinguish between a GW signal and the seismic background. As a result, the seismic noise of the GW measurement would be mitigated. However, so far, no quantitative assessment of the mitigation of the seismic background has been provided. In this article, we derive the analytical expressions for the optimal squared signal-to-noise ratio for a two-station array in an isotropic, random, Gaussian seismic field. Our numerical analysis reveals that the capacity to mitigate the seismic noise critically depends on the distance between the two stations relative to the seismic-correlation length. We demonstrate that optimal placement of the two stations can yield significant improvements in the equivalent seismic noise amplitude spectral density (ASD), approximately a factor of 2.3 at 0.3~Hz, compared to the measurement with a single station in the seismic-noise-dominated limit. The equivalent ASD of the seismic noise also exhibits distinct oscillatory and mitigation features arising from the Bessel-function structure of the noise correlation. 
\end{abstract}

\maketitle


\section{\label{sec:intro}Introduction }

The observation of gravitational waves (GWs) by the LIGO and Virgo detectors has inaugurated a new era of multi-messenger astronomy \cite{GW170817_MM}, allowing for unprecedented tests of general relativity \cite{GW170817_fund} and the exploration of compact binary populations \cite{GWTC3}. While current terrestrial detectors such as LIGO, Virgo, and KAGRA have achieved remarkable sensitivities in the high-frequency band ($f \gtrsim 10$\,Hz), they are fundamentally limited at lower frequencies by terrestrial seismic and gravitational noise (also known as Newtonian noise) \cite{1984PhRvD..30..732S, 2019LRR....22....6H, lowf_limits}. Space-based interferometers like LISA, TianQin and Taiji are poised to cover the milli-hertz regime \cite{2024arXiv240207571C,2016CQGra..33c5010L,2021PTEP.2021eA108L}. A sensitivity gap remains between a few tens of millihertz and a few Hz; a window crucial for multi-band observations and early warning alerts \cite{2016PhRvL.116w1102S,2020CQGra..37u5011A}. To access the rich physics in the deci-hertz band \cite{2025JCAP...01..108A}, which is populated by sources such as intermediate-mass black holes and the early inspiral of stellar-mass binaries, alternative detection concepts are required.

The Moon offers an exceptionally promising platform for filling this spectral gap. The lunar surface is characterized by a seismic background orders of magnitude quieter than Earth's, a lack of atmospheric disturbances, and cryogenic temperatures that naturally reduce thermal noise \cite{2009JGRE..11412003L,1999Icar..141..179V, nunn2020lunar}. Consequently, the concept of lunar GW detection has garnered significant interest \cite{2023SSRv..219...67B,2024RSPTA.38230066C}. By deploying a high-sensitivity seismometer array \cite{2014PhRvD..90j2001C,2021ApJ...910....1H} or a Laser interferometer \cite{2021CQGra..38l5008A,2025arXiv250915452P} on the lunar surface, one can exploit the Moon's low level of seismic vibrations to observe vibrations excited by passing GWs \cite{1960PhRv..117..306W,1968PhT....21d..34W,2009AdSpR..43..167P}.

However, the sensitivity of such a lunar-based observatory might be limited by the lunar seismic background \cite{seismic_moon_PRL}. Lunar GW detectors are mechanically coupled to the regolith, measuring the superposition of the GW-induced global deformation and the local seismic background. Previous studies estimating the sensitivity of lunar GW detectors have often simplified this noise model, assuming either single-station performance or uncorrelated noise averaged over an array \cite{2021ApJ...910....1H, 2024RSPTA.38230066C,2025arXiv250818437C}. 

Mitigation technologies based on seismic array data are under development for terrestrial GW detectors; in this case for the cancellation of Newtonian noise via Wiener filtering \cite{2016CQGra..33x4001C,Koley2024a}. Similarly, noise cancellation is being considered to reduce magnetic noise in GW detectors, which can be correlated between detectors across the globe \cite{2013PhRvD..87l3009T, 2018PhRvD..97j2007C}. Notably, in both examples, the data used to cancel the noise in GW detectors do not contain any (significant) GW signal. In the case of LGWA, the accelerometers contain a mix of GW signal and seismic background. Therefore, seismic mitigation for LGWA must be formulated as a signal-preserving coherent array filter rather than as standard witness-channel subtraction. As we will show explicitly in this paper, noise mitigation can still be achieved, but the calculation must now follow a different path from the noise-cancellation methods in terrestrial GW detectors. A common property of all these noise-cancellation schemes is that the optimal configuration of sensor arrays depends on the correlation length of the environmental field.

The LGWA array operates in a diffuse seismic background, which is expected to have a relatively short correlation length compared to typical terrestrial seismic fields  \cite{2024RSPTA.38230066C}. Not properly accounting for the spatial correlation of this background may lead to inaccurate estimations of the array's signal-to-noise ratio (SNR) with respect to a seismic background. The seismic background does not manifest as an uncorrelated instrument noise, but as a spatially correlated wavefield \cite{doi:10.1126/science.1078551,https://doi.org/10.1029/2005GL023518}.

In the deci-hertz band, the lunar seismic environment is distinct from that of Earth. The dominant contribution to the continuous seismic background is not tectonic activity or oceanic microseisms, but rather the stochastic ``hum" generated by the relentless flux of meteoroid impacts \cite{2009JGRE..11412003L, 2011Icar..211.1049G}. While the Moon also experiences transient and localized events, such as deep moonquakes (DMQs) that originate from specific nests ($700\text{--}1000~$km depth) with distinct periodicities \cite{2017JGRE..122.1487K, 2005JGRE..110.1001N}, and shallow thermal moonquakes driven by the diurnal cycle \cite{1974JGR....79.4351D,1979LPSC...10.2299N}, estimating the stationary sensitivity curve requires focusing on the continuous background floor. The meteoroid impact flux creates a diffuse, stationary surface-wave field that pervades the lunar crust \cite{2009JGRE..11412003L, 2012P&SS...74..179O}. Estimates of the continuous seismic background predict it to be orders of magnitude weaker than on Earth \cite{2009JGRE..11412003L}, but a spectral model of the background must still be provided, leaving important question marks about its strength in the deci-hertz band.

In this paper, we present a comprehensive theoretical framework for the mitigation of a random seismic background explicitly incorporating its spatial correlation structure. We model the meteoroid-induced noise as a stochastic field of fundamental-mode Rayleigh waves and analytically derive the optimal SNR for a two-sensor array. We demonstrate that the interplay between distance between stations and the correlation length of the seismic noise leads to frequency-dependent noise cancellation.

The paper is organized as follows. In Sec.~\ref{sec:theory}, we outline the theoretical framework, describing the tensor nature of the GW-induced lunar response and deriving the full covariance matrix for the seismic background using Bessel function expansions. In Sec.~\ref{sec:numerical_examples}, we present numerical examples for representative array configurations, analyzing the sky-averaged SNR and the resulting equivalent noise ASD curves. Finally, we summarize our findings and discuss implications for future array designs in Sec.~\ref{sec:con_dis}.

\section{Theoretical Framework}
\label{sec:theory}

To rigorously evaluate the noise-mitigation capability of a lunar seismometer network (like LGWA) for GW detection, we must construct a model that integrates the Moon's elastic response to GWs and the statistical properties of the lunar seismic-noise field. In the mid-frequency band ($0.01\sim1\,\text{Hz}$), the detection capability is partly constrained by the spatial coherence of the seismic background. In this section, we derive the optimal squared signal-to-noise ratio (SNR$^2$) for a two-seismometer array, accounting for the Moon's GW response and the correlation matrix of the Rayleigh-wave dominated background.

\subsection{Correlated Seismic Noise Field}
\label{subsec:noise_model}

The dominant environmental noise source for the lunar seismometer array is expected to be the seismic background generated by continuous micro-meteoroid impacts \cite{2009JGRE..11412003L,2021ApJ...910....1H}. We model this background as a stationary, stochastic surface-wave field. Assuming that meteoroid impacts are uniformly distributed over the lunar surface, the background in the mid-frequency band can be approximated as an isotropic field of Rayleigh waves \cite{doi:10.1126/science.1078551,2017PEPI..262...28G}.

Unlike simplified models that assume uncorrelated noise or scalar coherence, we adopt the complete form of the spatial autocorrelation (SPAC) matrix for multi-component observations. Following the formulation in previous works \cite{Aki_1957,1973ASAJ...54.1289C,2012GeoJI.191..189H}, for an isotropic field of Rayleigh waves, the cross-spectral density matrix $\bm{\Phi}(d, f)$ between two three-component seismometers separated by a horizontal distance $d$ is given by:
\begin{widetext}
\begin{equation}
\bm{\Phi}(d, f) = S_n^r(f) 
\begin{pmatrix} 
J_0(\zeta) & -\nu J_1(\zeta) & 0 \\ 
\nu J_1(\zeta) & \frac{\nu^2}{2}[J_0(\zeta) - J_2(\zeta)] & 0 \\ 
0 & 0 & \frac{\nu^2}{2}[J_0(\zeta) + J_2(\zeta)] 
\end{pmatrix}~,
\label{eq:haney_matrix}
\end{equation}
\end{widetext}
where $\zeta = 2\pi f d / c_R(f) = k_R(f) d$, where $c_R$ is the speed of a Rayleigh wave, and $J_n$ are Bessel functions of the first kind of order $n$. The matrix components correspond to the Vertical ($Z$), Radial ($R$, aligned with the separation vector \footnote{This is different from the radial unit vector $\hat{r}$ in spherical coordinates, which is introduced in Sec.~\ref{subsec:lunar_response}.}), and Transverse ($T$) directions. $S_n^r(f)$ is the power spectral density (PSD) of the vertical Rayleigh wave background (in units of m$^2 ~$Hz$^{-1}$), and $\nu(f)$ represents the horizontal-to-vertical (H/V) ratio (ellipticity) of the Rayleigh waves.

Equation (\ref{eq:haney_matrix}) reveals critical properties for array processing:
\begin{enumerate}
    \item \textbf{ZZ Correlation:} The vertical-vertical coherence follows $J_0(\zeta)$, which approaches unity for small separations ($d \to 0$) as $1 - \zeta^2/4$.
    \item \textbf{Horizontal Correlations:} The radial-radial ($RR$) and transverse-transverse ($TT$) correlations involve linear combinations of $J_0$ and $J_2$. Notably, even for isotropic noise, the coherence decay differs between longitudinal ($R$) and transverse ($T$) orientations.
    \item \textbf{Cross-Component Coupling:} The off-diagonal terms involving $J_1(\zeta)$ (i.e., $ZR$ or $RZ$) indicate non-zero correlation between vertical and horizontal components at non-zero distances, induced by the elliptical polarization of Rayleigh waves.
\end{enumerate}

The total noise covariance matrix $\mathbf{C}$ for a network also includes an incoherent instrumental noise component $P_{\text{inst},i}(f)$:
\begin{equation}
\mathbf{C}_{ij}(f) = P_{\text{inst},i}(f)\delta_{ij} + \hat{n}_i^T\mathbf{R}_{ij}^T \bm{\Phi}(|\vec{x}_i - \vec{x}_j|, f) \mathbf{R}_{ij} \hat{n}_j~,
\label{eq:noise-matrix}
\end{equation}
where $\mathbf{R}_{ij}$ is the rotation matrix aligning the global spherical coordinate system (introduced below) to the local Radial-Transverse frame defined by the pair $i,j$. $\hat{n}_i$ is the sensitive axis of i-th detector.

\subsection{Lunar Response and Spatial Derivatives}
\label{subsec:lunar_response}

Based on previous works \cite{1983NCimC...6...49B,2019PhRvD.100d4048M,PhysRevD.109.064092,4rrr-w4tm}, we adopt the updated spheroidal response formalism for the Moon. From now on we choose to use standard spherical coordinates centered at the Moon’s center, with three unit base vector fields $\hat{r}$, $\hat{\theta}$ and $\hat{\phi}$. For an incident monochromatic gravitational wave with strain tensor $\mathbf{h}$ (spatial component, in the TT gauge), the induced displacement field $\vec{\xi}(\vec{r})$ on the lunar surface ($r=R_{\text{M}}=$ lunar average radius) can be expressed in a compact, polarization-agnostic form \cite{2024PhRvD.110d3009Y}:
\begin{equation}
\vec{\xi}(\hat{r}) = 2 T_h \mathbf{h} \cdot \hat{r} + (T_r - 2T_h) (\hat{r} \cdot \mathbf{h} \cdot \hat{r}) \hat{r},
\label{eq:response_vector}
\end{equation}
where $T_r(f)$ and $T_h(f)$ are the frequency-dependent radial and horizontal response functions (per unit strain) determined by the Moon's radially heterogeneous structure (see \cite{PhysRevD.109.064092,2025PhRvD.111d4061M} for details).

Let $\hat{n}$ be the sensitive axis of a seismometer (e.g., $\hat{n} = \hat{r}$ for a vertical sensor). The signal measured at position $\vec{x}$ is therefore $s(\vec{x}) =\vec{\xi}(\vec{x}) \cdot \hat{n}$. For two nearby seismometers separated by $\vec{d}$, with baseline $\left | \vec{d} \right |  \ll R_{\text{M}}$, the variation in the GW response can be calculated as follows (for later convenience):

First, for two nearby sensors with different $\hat{n}$,
\begin{equation}
\delta s = \vec{\xi}(\hat{r}) \cdot \Delta\hat{n} ~.
\label{eq:signal_diff_approx1}
\end{equation}
If all the sensors are horizontal, we simply have
\begin{equation}
    \delta s \approx 2 T_h (\Delta \hat{n} \cdot \mathbf{h} \cdot \hat{r}) ~.
\end{equation}

Second, for two nearby sensors with similar $\hat{n}$, the difference in signal between them is well-approximated by the directional derivative:
\begin{equation}
\delta s(\vec{d},\hat{n}) \approx \vec{d} \cdot \nabla  [\vec{\xi}(\vec{r}) \cdot \hat{n}] ~.
\label{eq:signal_diff_approx}
\end{equation}
Furthermore, if both sensitive axes $\hat{n}$ are strictly aligned with one of the coordinate axes ($\hat{r}$, $\hat{\theta}$ or $\hat{\phi}$), $\delta s$ can be further calculated as
\begin{align}
    \delta s(\vec{d},\hat{r}) \approx & \frac{2 T_r}{R_{\text{M}}} ( h_{\theta r} d_{\theta} + h_{\phi r} d_{\phi}) \nonumber \\
    \delta s(\vec{d},\hat{\theta}) \approx & \frac{2 T_h}{R_{\text{M}}} [(h_{\theta \theta}-h_{rr})d_{\theta} + (h_{\theta \phi}+h_{\phi r} \cot \theta)d_{\phi}] \nonumber \\
    \delta s(\vec{d},\hat{\phi}) \approx & \frac{2 T_h}{R_{\text{M}}} [h_{\theta \phi}d_{\theta} + (h_{\phi \phi}-h_{r r} - h_{\theta r}\cot \theta)d_{\phi}]~.
    \label{eq:delta_s_align}
\end{align}

\subsection{SNR$^2$ Density for a two-sensor Array}

For simplicity, from now on we assume identical instrument noise PSDs of all seismic channels, i.e., $P_{\text{inst},i} = P_{\text{inst}}$, and rewrite the noise matrix for two sensors as follows:
\begin{align}
    \mathbf{C} = \begin{bmatrix}
 P_{\text{inst}} + \alpha_1 S_n^r & \gamma S_n^r\\ 
 \gamma S_n^r & P_{\text{inst}} + \alpha_2 S_n^r
\end{bmatrix} ~,
\end{align}
in which
\begin{align}
    \alpha_i = &1-n_{h,i}^2 + \frac{\nu^2}{2} n_{h,i}^2 \nonumber \\
    \gamma = & \hat{n}_1^T\mathbf{R}_{12}^T \bm{\Phi}(d, f) \mathbf{R}_{12} \hat{n}_2 / S_n^r ~,
\end{align}
and $n_{h,i}$ is the horizontal component of $\hat{n}_i$. The inverse matrix can then be calculated as
\begin{align}
    \mathbf{C}^{-1} = \frac{1}{\Delta} \begin{bmatrix}
 P_{\text{inst}} + \alpha_2 S_n^r & -\gamma S_n^r\\ 
 -\gamma S_n^r & P_{\text{inst}} + \alpha_1 S_n^r
\end{bmatrix} ~,
\end{align}
in which
\begin{equation}
    \Delta = (P_{\text{inst}} + \alpha_1 S_n^r)(P_{\text{inst}} + \alpha_2 S_n^r) - (\gamma S_n^r)^2 ~.
\end{equation}
For strictly horizontal sensors, we have
\begin{align}
    \alpha = &\frac{\nu^2}{2} \nonumber \\
    \gamma = & \frac{\nu^2}{2} [J_0 (k_R d) \cos (\tau_1 - \tau_2) \nonumber \\
    &- J_2 (k_R d) \cos (\tau_1 + \tau_2)] ~,
\end{align}
in which $\tau_i$ is the angle between $\hat{n}_i$ and $\vec{d}$.

The SNR$^2$ density $\mathcal{R}(f)$ for multiple detectors with signal vector $\mathbf{s} = (s_1, s_2, ...)^T$ and covariance matrix $\mathbf{C}$ is \cite{2017LRR....20....2R}:
\begin{equation}
\mathcal{R} = \mathbf{s}^\dagger \mathbf{C}^{-1} \mathbf{s}.
\end{equation}
Although this equation formally represents the optimal expected SNR$^2$ for a known signal, it also corresponds to a physical array-filtering operation: the data are combined with inverse-covariance weights while preserving the expected GW response \cite{PhysRevD.64.042004,2005CQGra..22S1321W}. This coherent spatial filter differs from terrestrial Wiener subtraction, which relies on signal-free witness channels \cite{2016CQGra..33x4001C,Koley2024a}, because the LGWA seismic channels themselves contain the GW signal.

For two sensors, we can decompose this into common ($s_+ = s_1+s_2$) and differential ($\delta s = s_2-s_1$) modes, and assume horizontal sensors. We then obtain:
\begin{equation}
\mathcal{R} = \frac{1}{2} \bigg[\frac{|s_+|^2}{P_{\text{inst}} + \nu^2 S_n^r (1+\Gamma)/2} + \frac{|\delta s|^2}{P_{\text{inst}} + \nu^2 S_n^r (1-\Gamma)/2} \bigg] ~,
\label{eq:snr_modes}
\end{equation}
where $\Gamma \equiv \gamma/\alpha$ is the normalized correlation function. This decomposition shows explicitly that the optimal coherent filter weights the common and differential modes according to their noise variances \footnote{For comparison, a signal-free witness channel would leave the standard Wiener residual proportional to $1-\Gamma^2$, whereas a signal-preserving combination of two channels with equal GW responses leaves a residual proportional to $(1+\Gamma)/2$.}. When instrument noise dominates, i.e., $\nu^2 S_n^r \ll P_{\text{inst}}$, we simply recover the standard result
\begin{equation}
\mathcal{R}_{\text{ins}} = \frac{s_1^2+s_2^2}{P_{\text{inst}}} ~.
\end{equation}
When the environmental noise completely dominates, i.e., $(1-\Gamma) \gg P_{\text{inst}}/(\nu^2 S_n^r)$, we have
\begin{align}
\mathcal{R}_{\text{env}} = &\frac{1}{\nu^2 S_n^r} \bigg(\frac{|s_+|^2}{1+\Gamma} + \frac{|\delta s|^2}{1-\Gamma} \bigg) ~.
\end{align}

\subsection{Optimal Filtering for Two Close-Proximity Horizontal Sensors}
\label{subsec:two_detector}

We now derive the optimal SNR$^2$ density for a pair of horizontal sensors separated by a short distance $d$ ($k_R d \ll 1$), i.e.,
\begin{equation}
    d \ll 1.6~\text{km} \left ( \frac{c_R}{1000~\text{m/s}}  \right ) \left ( \frac{f}{0.1~\text{Hz}}  \right ) ^{-1} ~.
    \label{eq:close-proximity-criteria}
\end{equation}
In this case, the correlation factor can be approximated as
\begin{align}
    \Gamma \simeq & \cos (\tau_1 - \tau_2) \nonumber \\
    &- \frac{(k_R d)^2}{8}[2\cos (\tau_1 - \tau_2)+\cos (\tau_1 + \tau_2)] \nonumber \\
    &+ \mathcal{O}[(k_R d)^4]~.
\end{align}

\subsubsection{Parallel detectors ($\tau_1 = \tau_2$)}

When the sensitive axes of two sensors are perfectly parallel, the normalized correlation function can be approximated as
\begin{equation}
    \Gamma^{\text{Par}} \simeq 1 - \frac{(k_R d)^2}{8} (2 + \cos 2\tau) ~,
\end{equation}
and the SNR$^2$ density are
\begin{align}
    \mathcal{R}^{\text{Par}} \simeq & \frac{1}{2} \bigg[\frac{|s_+|^2}{P_{\text{inst}} + \nu^2 S_n^r } \nonumber\\ 
    & + \frac{|\delta s|^2}{P_{\text{inst}} + \nu^2 S_n^r (k_R d)^2 (2 + \cos 2\tau)/16} \bigg]\\
    \mathcal{R}_{\text{env}}^{\text{Par}} \simeq &\frac{1}{\nu^2 S_n^r} \bigg[\frac{|s_+|^2}{2} + \frac{8 |\delta s|^2}{(k_R d)^2 (2 + \cos 2\tau)} \bigg] ~.
    \label{eq:SNR_env_Par}
\end{align}
for ordinary and environmental-noise-dominant cases respectively. A special situation occurs when $\nu^2 S_n^r (k_R d)^2 \ll P_{\text{inst}} \ll \nu^2 S_n^r$:
\begin{equation}
    \mathcal{R}_{\text{env-inst}}^{\text{Par}} \simeq \frac{1}{2} \bigg(\frac{|s_+|^2}{ \nu^2 S_n^r } + \frac{|\delta s|^2}{P_{\text{inst}} } \bigg) ~.
    \label{eq:SNR_env-inst_Par}
\end{equation}

We notice that, in the parallel case $\delta s /s \sim \mathcal{O} (d/R_{\text{M}})$, as seen in Eq.~(\ref{eq:delta_s_align}). Hence we observe that the first term in Eq.~(\ref{eq:SNR_env_Par}) almost always dominates, because the opposite requires
\begin{align}
    &\frac{|\delta s|^2}{(k_R d)^2} \ge |s|^2~, \nonumber \\
    \iff &~ k_R R_{\text{M}} \le 1~, \nonumber \\
    \iff &~ f \le 0.18~\text{mHz}\left ( \frac{c_R}{2000~\text{m/s}}  \right ) ~,
\end{align}
which is far below the target frequency band of lunar seismic GW detection. For the same reason, the first term in Eq.~(\ref{eq:SNR_env-inst_Par}) will also always dominate.

\subsubsection{Non-parallel detectors ($\tau_1 \ne \tau_2$)}

The SNR$^2$ density for two non-parallel sensors is
\begin{align}
    \mathcal{R}_{\text{env}}^{\text{NP}} &\simeq \frac{1}{\nu^2 S_n^r} \bigg[\frac{|s_+|^2}{1+\cos (\tau_1 - \tau_2)} + \frac{|\delta s|^2}{1-\cos (\tau_1 - \tau_2)} \bigg] \nonumber \\
    &= \frac{2}{\nu^2 S_n^r} \frac{\delta s^2+2s_1 s_2 [1-\cos (\tau_1 - \tau_2)]}{\sin^2 (\tau_1 - \tau_2)}~.
\end{align}

\quad

\subsection{Strain Noise Estimation and Equivalent Noise Spectrum}

Based on the calculations of SNR$^2$ density, we can directly calculate the (dimensionless) strain noise of the detector array completely induced by environmental noise:
\begin{align}
    h_{c,n} =& \sqrt{f}\frac{h_0}{\sqrt{\mathcal{R}(f)}} \nonumber \\
    \simeq&  \nu \frac{\sqrt{f S_n^r}}{T_{r/h}} \times \mathcal{M} (\hat{r},\hat{n},\mathbf{h})~,
    \label{eq:strain}
\end{align}
where we assume that the signals are induced by GW with amplitude $h_0 (f)$ (in units of $\text{Hz}^{-1}$). $\mathcal{M} (\hat{r},\hat{n},\mathbf{h})$ is an $\mathcal{O}(1)$ mitigation factor depending on the array configuration and GW polarization. Our sensitivity estimation aims to define an effective sky-averaged noise floor rather than the detectability of a specific source. By calculating the optimal SNR for a generic deterministic GW signal, we will establish an equivalent strain noise ASD [Eqs.~(\ref{eq:ASD}) and (\ref{eq:averaged_mitigation})], which can be regarded as a standardized baseline to evaluate the environmental noise limit for any modeled GW searches.

Assuming that $T_{r/h} \simeq L(f) \times 10^6~\text{m}$ ($L$ is dimensionless), and the amplitude spectral density of Rayleigh wave background caused by meteoroid impact is approximately \cite{2009JGRE..11412003L,2021ApJ...910....1H,2024RSPTA.38230066C}:
\begin{equation}
    \sqrt{S_n^r} \simeq 10^{-12\;\text{to}\;-16} \left ( \frac{f}{0.1~\text{Hz}}  \right )^{-\beta} ~\text{m Hz}^{-1/2} ~
\end{equation}
with $0< \beta <2 $. As a result, we have
\begin{align}
    h_{c,n} \simeq  3.16\times10^{-19\;\text{to}\;-23} \frac{\nu}{L (f)} \left ( \frac{f}{0.1~\text{Hz}}  \right )^{-\beta+0.5} ~.
\end{align}

To directly compare the environmental limits with instrument noise, and to eliminate interference from lunar response ambiguity, we will focus on the equivalent displacement ASD of seismic noise in the following numerical analysis, which is defined as:
\begin{equation}
    A_s (f) \equiv \nu \mathcal{M}\sqrt{S_n^r} ~.
    \label{eq:ASD}
\end{equation}

\section{Numerical Examples for Detector Array}
\label{sec:numerical_examples}

\begin{figure}
\includegraphics[width=0.98\linewidth]{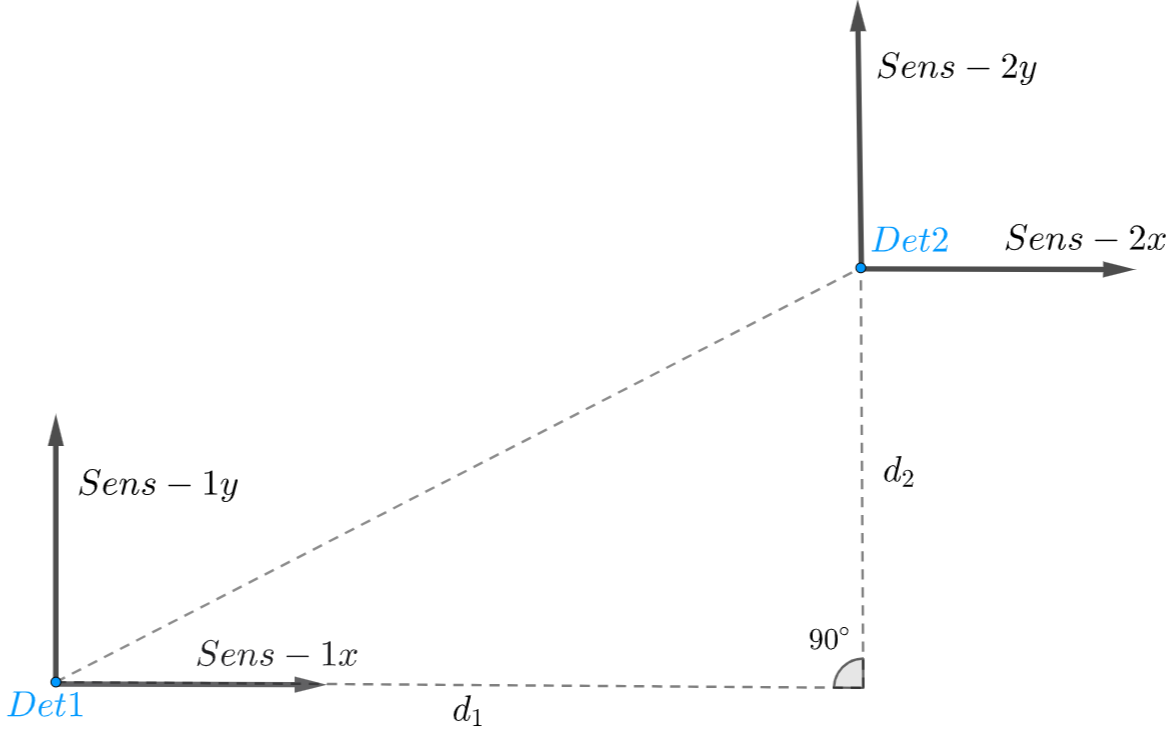}
\caption{Array configuration scenario.}
\label{fig:fig1}
\end{figure}

In this section, we calculate the sky-averaged SNR$^2$ density, defined as
\begin{equation}
    \left \langle \mathcal{R} \right \rangle \equiv  \frac{1}{4 \pi^2} \int \mathcal{R} ~\mathrm{d}\Omega ~\mathrm{d}\psi ~,
\end{equation}
in which $\Omega$ describes the GW propagation direction (including two angles in standard spherical coordinates), and $\psi$ is the GW polarization angle (see the details in e.g., \citet{2019PhRvD.100d4048M}).
We set a simple but representative array configuration, shown in Fig.~\ref{fig:fig1}. Motivated by the practical engineering limitation that high-sensitivity lunar seismometers (such as the proposed LGWA) are optimized for horizontal measurements \cite{2025JCAP...01..108A}, we assume there are two horizontal detectors (Det1 and Det2) and each of them has two mutually perpendicular sensors. Their separations, $d_1$ and $d_2$, are much smaller than the lunar radius $R_{\text{M}}$, but not necessarily satisfying Eq.~(\ref{eq:close-proximity-criteria}). We assume that the noise matrix for all the four sensors can be described by Eq.~(\ref{eq:noise-matrix}), while we also assume in the following calculations that $S_n^r \gg P_{\text{inst}}$ in order to isolate the effect of seismic noise.

\begin{figure}
\includegraphics[width=0.98\linewidth]{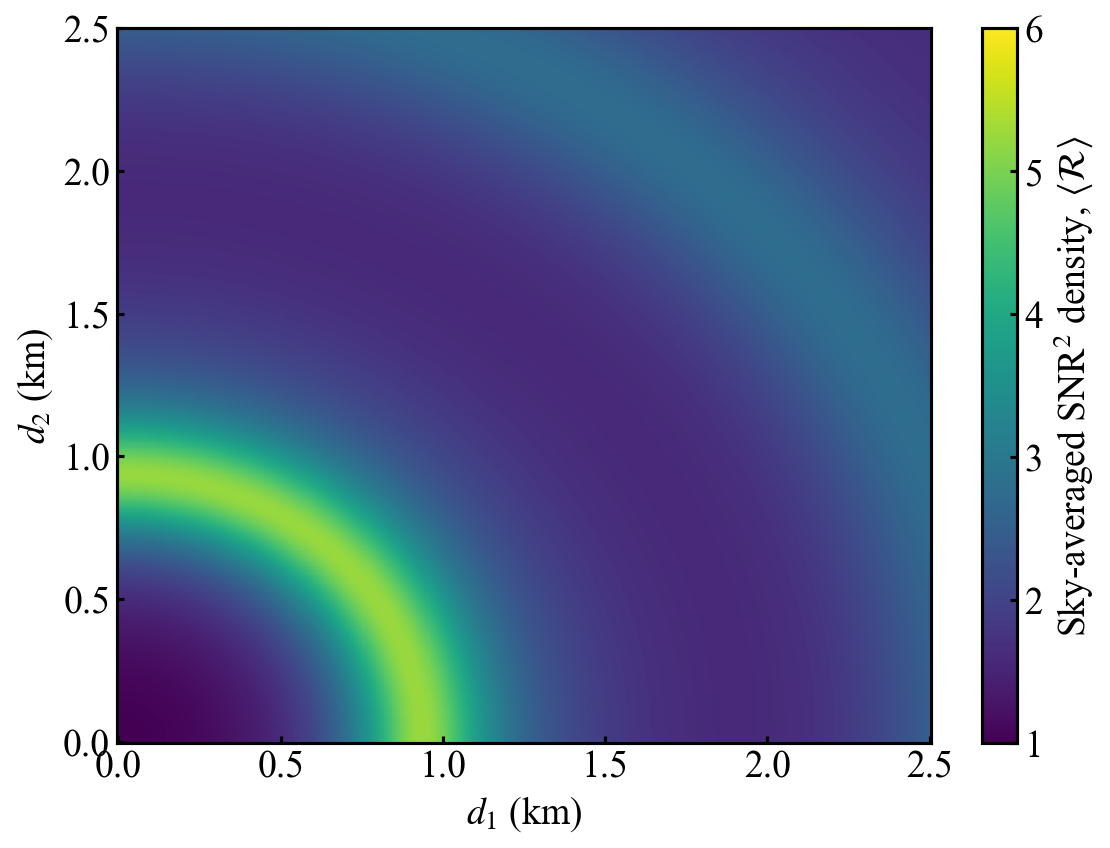}
\caption{Sky-averaged SNR$^2$ density, normalized by zero-separation ($d_1=d_2=0$) value. $f=0.3~\text{Hz}$ and $c_R=500~\text{m/s}$.}
\label{fig:fig2}
\end{figure}

We first calculate $\left \langle \mathcal{R} \right \rangle$ for $f=0.3~\text{Hz}$ (optimal frequency for LGWA \cite{2025JCAP...01..108A}) and $c_R=500~\text{m/s}$ (which means $k_R = 3.77\times 10^{-3} ~\text{m}^{-1}$) in Fig.~\ref{fig:fig2}. We normalize the results by zero-separation ($d_1=d_2=0$) value, in order to isolate the array configuration and environmental effect from GW-waveform features. We also set $S_n^r/P_{\text{inst}}=100$ and $\nu=0.5$ to investigate the environment-dominant case. It is worth noting that in the strictly seismic-noise-dominated regime considered here, a zero-separation two-station array measures identical, perfectly correlated seismic noise. Consequently, its intrinsic seismic-noise floor is mathematically equivalent to that of a single station. The results reflect a clear distance-dependent and direction-independent behavior. The maximum value ($\sim 5.2$) around $\sqrt{d_1^2+d_2^2}\simeq 0.93~\text{km}$ corresponds to $\zeta=k_R d\simeq 3.5$. These approximately correspond to the behavior of 
\begin{equation}
    \frac{2}{1+J_0 (\zeta)-0.5J_2 (\zeta)} ~.\nonumber
\end{equation}
We point out here that the above formula is just a phenomenological approximation of the numerical results. To demonstrate its accuracy, we compare the exact numerical values with this analytical approximation in Appendix~\ref{app:approxtest}. The comparison indicates that the approximation remains accurate within a relative error of $25\%$ across the relevant $\zeta$ range. While this specific approximation applies to the two-station (with four-sensor) scenario, our theoretical formulation is fully generalizable to multi-detector arrays.
Our calculations (not plotted here) also show that, for this two-station array, the orientation of the second detector relative to the first one does not influence the sky-averaged SNR$^2$ density \footnote{This is incorrect for GW with single polarization and wave-direction, as shown in Appendix~\ref{app:sing-pol}.}. This is understandable because a detector with two mutually perpendicular sensors can in principle extract all the information from horizontal signal. For reference, we also calculate the results for a two-sensor array (Sens-1x and Sens-2x; Sens-1x and Sens-2y) in Appendix~\ref{app:two-sensor}, and calculate the results for single-polarization GW in Appendix~\ref{app:sing-pol}.

\begin{figure}
\includegraphics[width=0.98\linewidth]{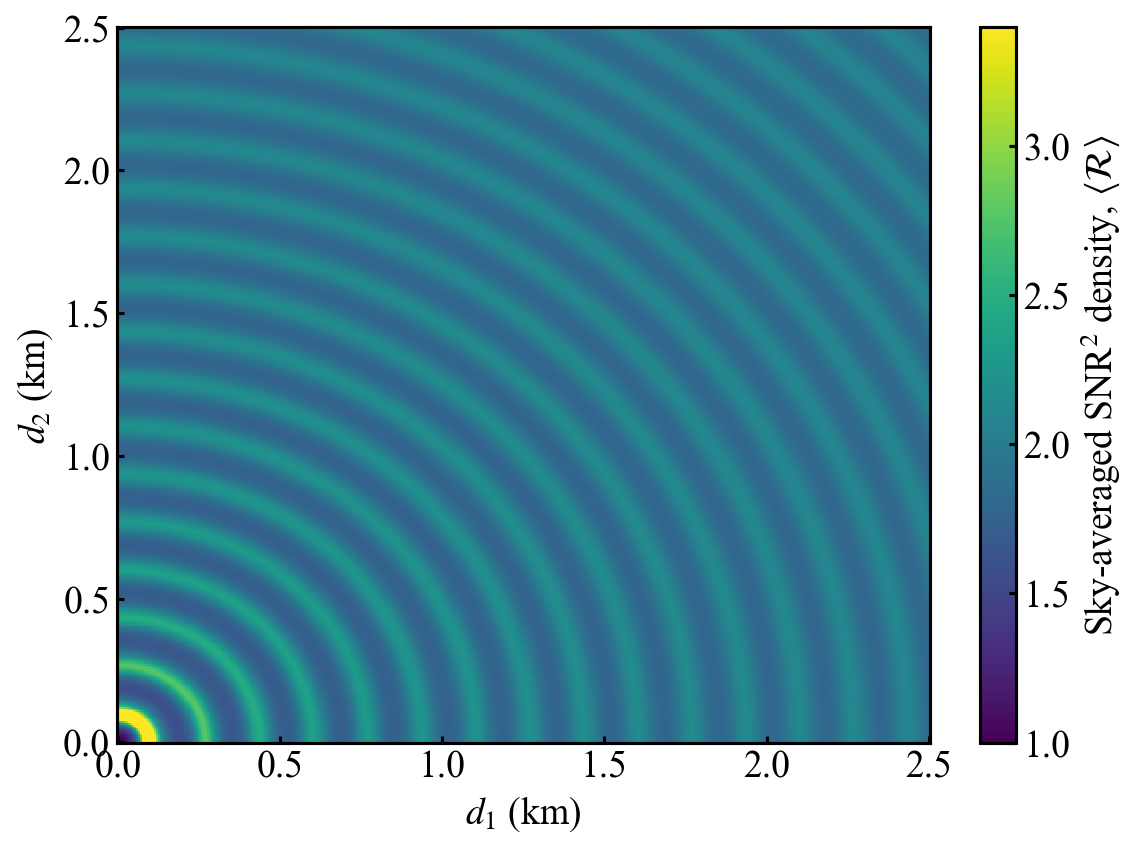}
\caption{Sky-averaged SNR$^2$ density (short-coherence case), normalized by zero-separation ($d_1=d_2=0$) value. $f=3~\text{Hz}$ and $c_R=500~\text{m/s}$.}
\label{fig:fig3}
\end{figure}

We next calculate the short-coherence case, which means that we should let $k_R d \gg 1$. We fulfill this condition by simply changing the frequency to $f = 3~\text{Hz}$ and keeping other parameters unchanged. The results are shown in Fig.~\ref{fig:fig3}. We find a 2 times gain in the sky-averaged SNR$^2$ density at large separation on average, and also find an obvious spatial period of $\Lambda\sim 170~\text{m}$ at large distance, corresponding to $k_R \Lambda \sim 2\pi$.

\begin{figure}
\includegraphics[width=0.98\linewidth]{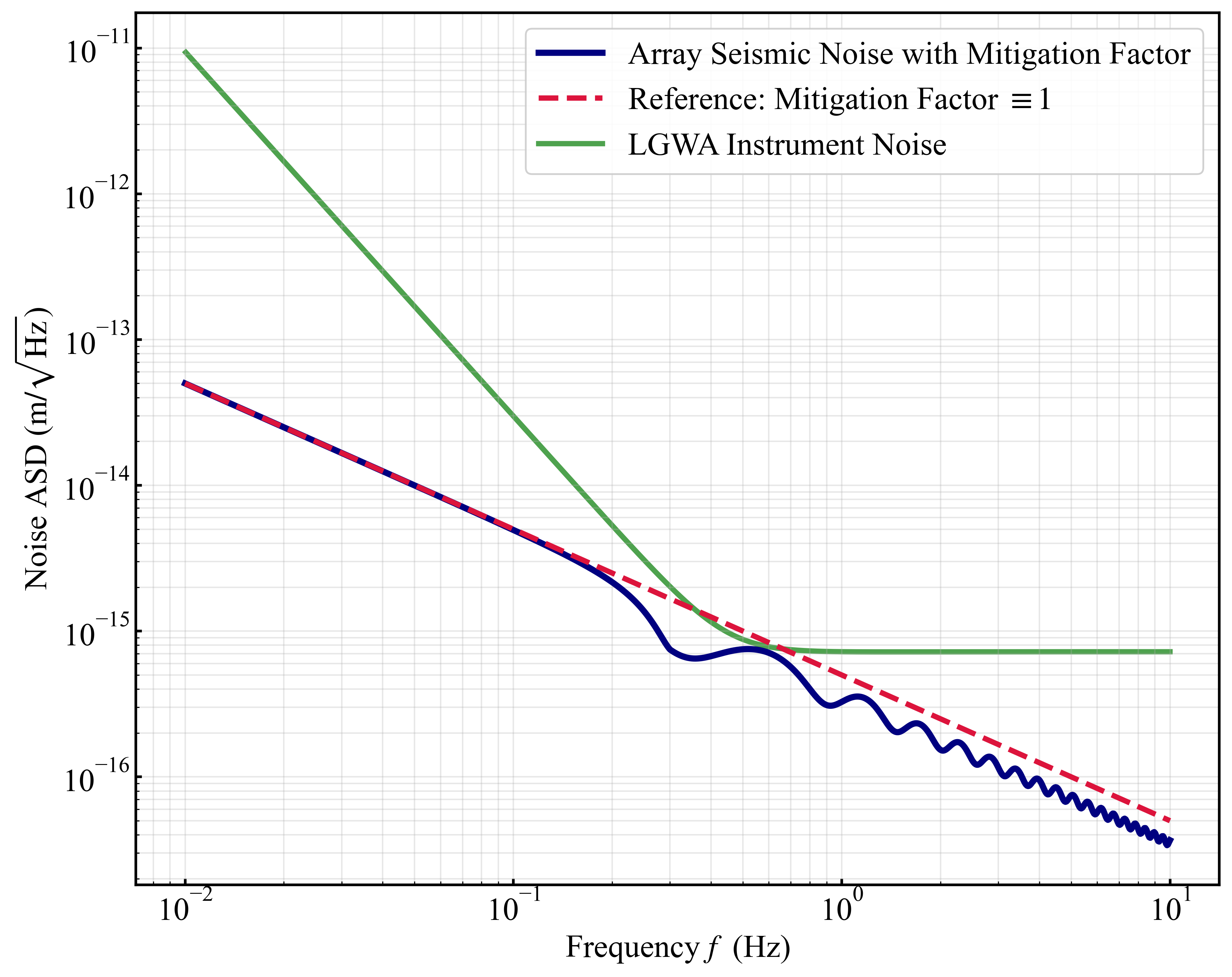}
\caption{
Equivalent noise ASD of the seismic background without and with array mitigation in the seismic-noise-dominated limit. The detector separation is fixed as 900\,m. The green curve shows the nominal instrument noise model adopted here for one horizontal accelerometer \cite{webs}. It contains no array averaging or seismic-noise mitigation and is shown only as a reference.}
\label{fig:fig4}
\end{figure}

Finally, based on previous calculations and discussions, we can now give a better description to the behavior of equivalent seismic noise ASD [Eq.~(\ref{eq:ASD})]. We fix the background spectrum as (with $\nu=0.5$)
\begin{equation}
    \sqrt{S_n^r} \simeq 10^{-14} \left ( \frac{f}{0.1~\text{Hz}}  \right )^{-1} ~\text{m Hz}^{-1/2} ~,
\end{equation}
and assume the wave velocity behaves like 
\begin{align}
    &c_R [\text{km/s}]\simeq \nonumber\\ &\left\{\begin{matrix}
 2,~f\le 50~\text{mHz} \\[0.15cm]
2-1.93~\text{log}_{10}(\frac{f}{50~\text{mHz}} ) ,~ 50~\text{mHz}< f<0.3~\text{Hz} \\[0.2cm]
0.5,~f\ge 0.3~\text{Hz}
\end{matrix}\right. ~.
\label{eq:wavespeed}
\end{align}
This is a phenomenological formula trying to include both the shallow regolith properties and low-frequency characteristic induced by large-scale structure \cite{2011Sci...331..309W}, largely similar to the surface wave velocity calculated in \citet{2014PhRvD..90j2001C}.
Finally, we select an effective mitigation factor:

\begin{align}
    \mathcal{M}_{\text{eff}}  &\equiv \sqrt{\frac{\left \langle \mathcal{R} \right \rangle _{d=0} }{\left \langle \mathcal{R} \right \rangle} }  \nonumber \\
    &\sim \sqrt{1+J_0 (\zeta)-0.5 J_2(\zeta)}/\sqrt{2} ~,
    \label{eq:averaged_mitigation}
\end{align}
and fix the array separation as $d=0.9~$km. Here the mitigation factor $\mathcal{M}_{\text{eff}}$ is defined as the residual ASD factor of the optimal coherent filter relative to the zero-separation baseline. Equivalently, $1/\mathcal{M}_{\text{eff}}$ is the ASD reduction factor.
We calculate the noise ASD [Eq.~(\ref{eq:ASD})] with ($\mathcal{M} = \mathcal{M}_{\text{eff}}$) and without $\mathcal{M}_{\text{eff}}$ ($\mathcal{M} = 1$), and plot the results in Fig.~\ref{fig:fig4}. Note that the equivalent seismic noise ASD we plot here describes the strictly seismic-noise-dominated case \footnote{Although we add an instrument noise curve in Fig.~\ref{fig:fig4}, it is just for comparison and is not considered in the calculations in this work.}. If instrumental noise is comparable to or exceeds the seismic background, the full covariance matrix [Eq.~(\ref{eq:noise-matrix})] must be included in the pipeline, and the overall mitigation efficacy would inherently depend on the frequency-dependent ratio $S_n^r(f)/P_{\text{inst}}(f)$. The oscillations observed at higher frequencies arise from the Bessel function structure of the noise correlation, reflecting the interplay between the array baseline and the frequency-dependent seismic wavelength. Due to this effect, there is an approximately 2.3 times mitigation of the ASD at frequencies near 0.3 Hz, and an approximately $\sqrt{2}$ times mitigation in the high-frequency limit.

\section{Conclusions and Discussions}
\label{sec:con_dis}

In this paper, we present a comprehensive theoretical framework to evaluate the noise-mitigation capability of a lunar seismometer array optimized for the detection of GWs. By integrating the Moon's global elastic response to GWs with the spatial autocorrelation model of the meteoroid-induced seismic noise, we derived the analytical expression for the squared signal-to-noise ratio (SNR$^2$) density of a two-sensor array. This framework allows for a rigorous assessment of how array geometry and environmental noise properties jointly constrain the detectability of GW signals in the mid-frequency band. While our numerical analysis demonstrates this using a representative two-station configuration, the analytical formulation is intrinsically scalable to networks comprising an arbitrary number of sensors.

Our numerical analysis reveals that the noise mitigation of the lunar array depends on the array configuration relative to the correlation length of the seismic field. As illustrated in Fig.~\ref{fig:fig2} and Fig.~\ref{fig:fig3}, proper array design can yield significant improvements in the SNR to GWs compared to the zero-separation baseline. Specifically, we observed an enhancement factor of approximately $\sqrt{5.2} \simeq 2.3$ in the low-frequency regime ($f=0.3$\,Hz) and $\sqrt{2} \simeq 1.4$ in the higher-frequency regime ($f=3$\,Hz). These gains arise from the distinct spatial coherence signatures of the GW signal, which induces global, coherent deformations, versus the seismic noise, which de-correlates or exhibits specific phase relationships (described by Bessel functions) over shorter spatial scales.

Extending this analysis to the broadband regime, Fig.~\ref{fig:fig4} illustrates the array's equivalent noise ASD. The noise-cancellation efficacy is inherently frequency-dependent due to the Bessel-function structure of the spatial coherence. By incorporating a phenomenological dispersion model [Eq.~(\ref{eq:wavespeed})] for realistic evaluation, we observe distinct oscillatory features as the varying effective array aperture ($d/\Lambda$) modulates the residual noise floor. Our calculation reveals the intrinsic limits of such array-based mitigation: the suppression of an isotropic seismic background is generally constrained to within a factor 3. Furthermore, because the ratio of the array baseline to the seismic wavelength varies across the observation band, an effect further compounded by the dispersive nature of $c_R(f)$ in the lunar regolith, a fixed array geometry cannot provide optimal mitigation across all frequencies. 

While prior sensitivity studies (like \cite{2024RSPTA.38230066C}) typically estimated the noise floor without explicitly accounting for this coherent spatial cancellation, our results demonstrate that in the case of two seismometers, seismic noise from an isotropic field cannot be mitigated by large factors. Although this background does not drastically compromise the target sensitivity of LGWA, it necessitates that future-array designs perform a multi-objective optimization, convoluting the lunar GW response with the seismic dispersion curve of the deployment site.

The robustness of our results relies on the accuracy of our model of the lunar seismic wave-field. Our current model assumes an isotropic incidence of fundamental-mode Rayleigh waves. In reality, a certain level of anisotropy, non-negligible body-wave content, and non-stationarity of the field are to be expected \cite{2010EaSci..23..519S}. The scattering of waves by the heterogeneous megaregolith may induce frequency-dependent anisotropy, particularly if the meteoroid impact distribution is non-uniform or if local topography induces scattering and focusing \cite{2012JGRE..117.6003B,https://doi.org/10.1029/2020JE006406}. The isotropic wavefield adopted here provides a conservative reference case for the diffuse, sky-averaged configurations considered in this work, as it contains no preferred direction that can be exploited by array processing. If the background deviates from this assumption, the spatial covariance matrix $\Phi(d,f)$ would encode distinct directional and phase information. Depending on the directional covariance, phase velocities, and the overlap between the dominant noise eigenmodes and the GW response, the optimal inverse-covariance filter may exploit such spatial structures and achieve higher mitigation efficiencies. However, a universal improvement cannot be assumed without specifying a particular non-isotropic covariance model. The LGWA array is planned with four seismic stations \cite{2025JCAP...01..108A} (discussed in Appendix~\ref{app:four-station}) to achieve effective noise mitigation for varying levels of anisotropy, for which both the inter-station distances and the array configuration, including baseline orientations, must be jointly optimized to maximize noise mitigation in the most affected frequency band.

In conclusion, our work establishes the first preliminary framework bridging the gap between phenomenological models of lunar surface wave scattering and the noise budgeting of lunar-based GW detectors. It highlights the profound synergy between GW astronomy and lunar geophysics: while seismic noise is a nuisance for GW detection, its precise characterization is a scientific objective for understanding the Moon's shallow structure. Future studies should extend this framework to the full LGWA array, including body-wave contributions at higher frequencies, anisotropic noise fields, and the development of adaptive data processing algorithms to optimize mitigation of non-stationary seismic fields.

\begin{acknowledgments}
We thank Xian Chen for many helpful discussions. This work is supported by the National Key Research and Development Program of China (Grant No.~2024YFC2207300) and the Italian Space Agency (ASI) under Grant No.~2025-29-HH.0. Han Yan acknowledges support from the China Scholarship Council (No.~202506010256).
\end{acknowledgments}

\appendix
\setcounter{figure}{0}
\renewcommand{\thefigure}{A\arabic{figure}}

\section{Sky-averaged SNR$^2$ density for a two-sensor array}
\label{app:two-sensor}

\begin{figure}[htbp] 
    \centering
    
    \begin{subfigure}{}
        \includegraphics[width=\linewidth]{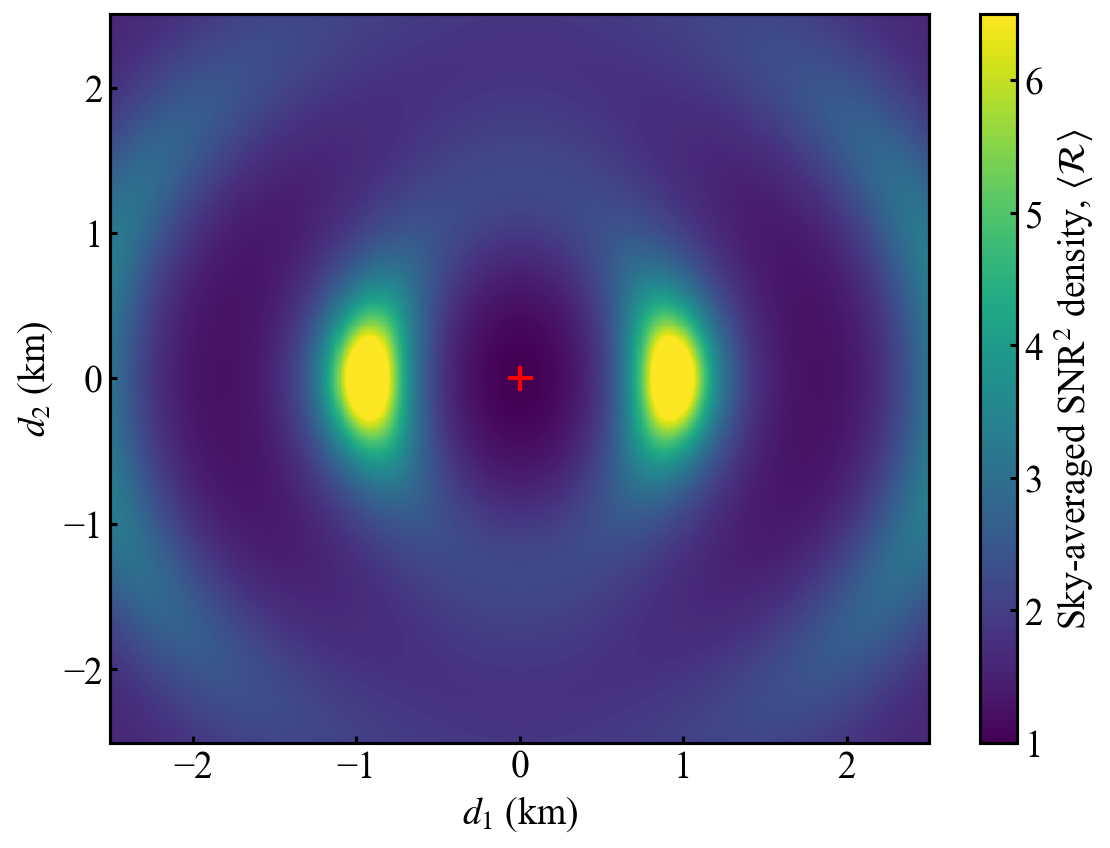}
    \end{subfigure}
    \begin{subfigure}{}
        \includegraphics[width=\linewidth]{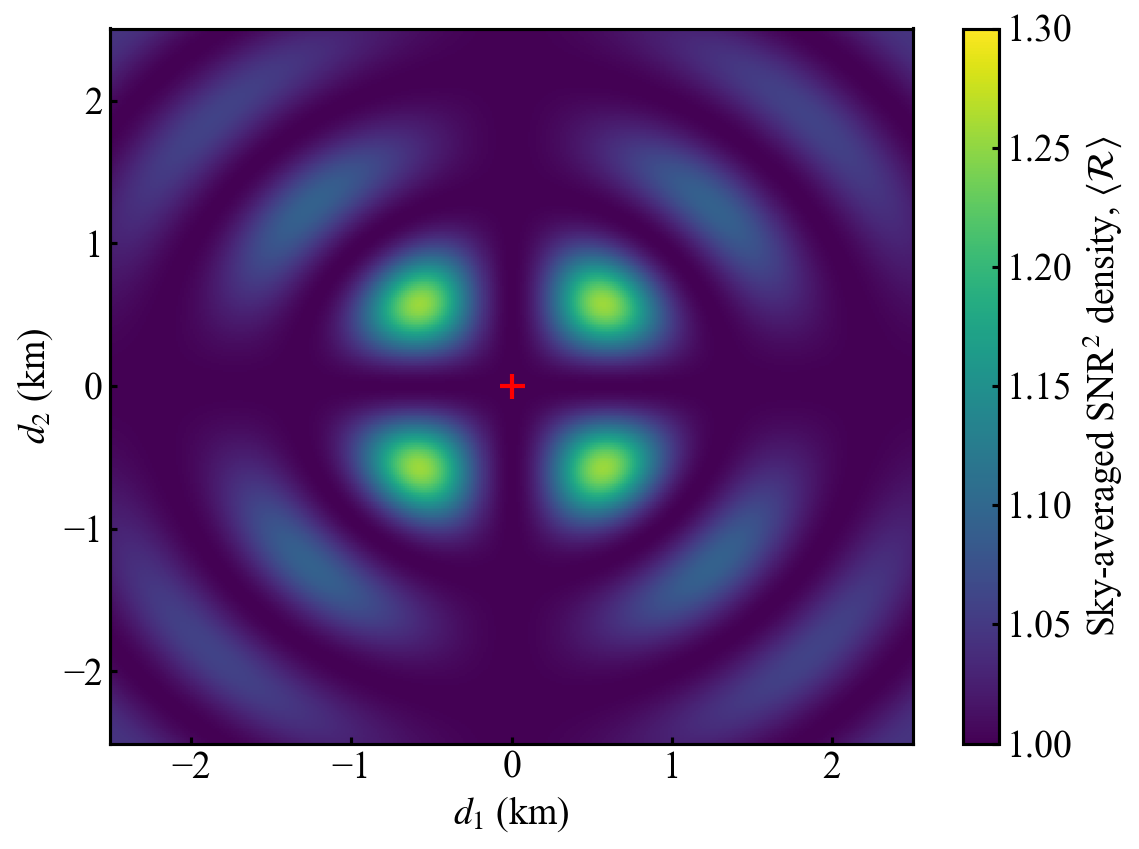}
    \end{subfigure}
    
    \caption{Sky-averaged SNR$^2$ density for a two-sensor array. Upper panel: Sens-1x and Sens-2x array; Lower panel: Sens-1x and Sens-2y array. Both panels are normalized by zero-separation ($d_1=d_2=0$, marked with red cross) value. $f=0.3~\text{Hz}$ and $c_R=500~\text{m/s}$.}
    \label{fig:2det}
    
\end{figure}

In this Appendix, we plot the sky-averaged SNR$^2$ density for a two-sensor array in Fig.~\ref{fig:2det}: Sens-1x with Sens-2x in the upper panel; Sens-1x with Sens-2y in the lower panel. The results match well with our analytical formulations.

\section{SNR$^2$ density for single polarization}
\label{app:sing-pol}

\begin{figure}[htbp] 
    \centering
    
    \begin{subfigure}{}
        \includegraphics[width=\linewidth]{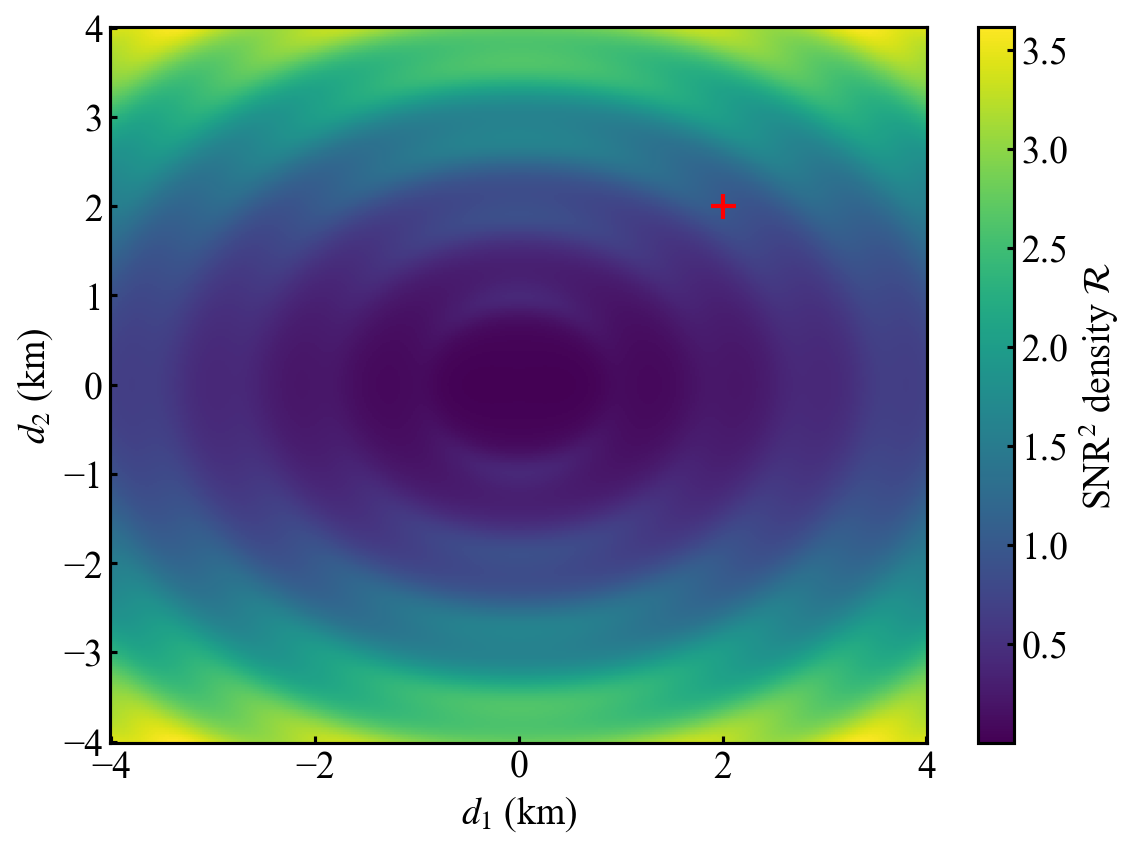}
    \end{subfigure}
    \begin{subfigure}{}
        \includegraphics[width=\linewidth]{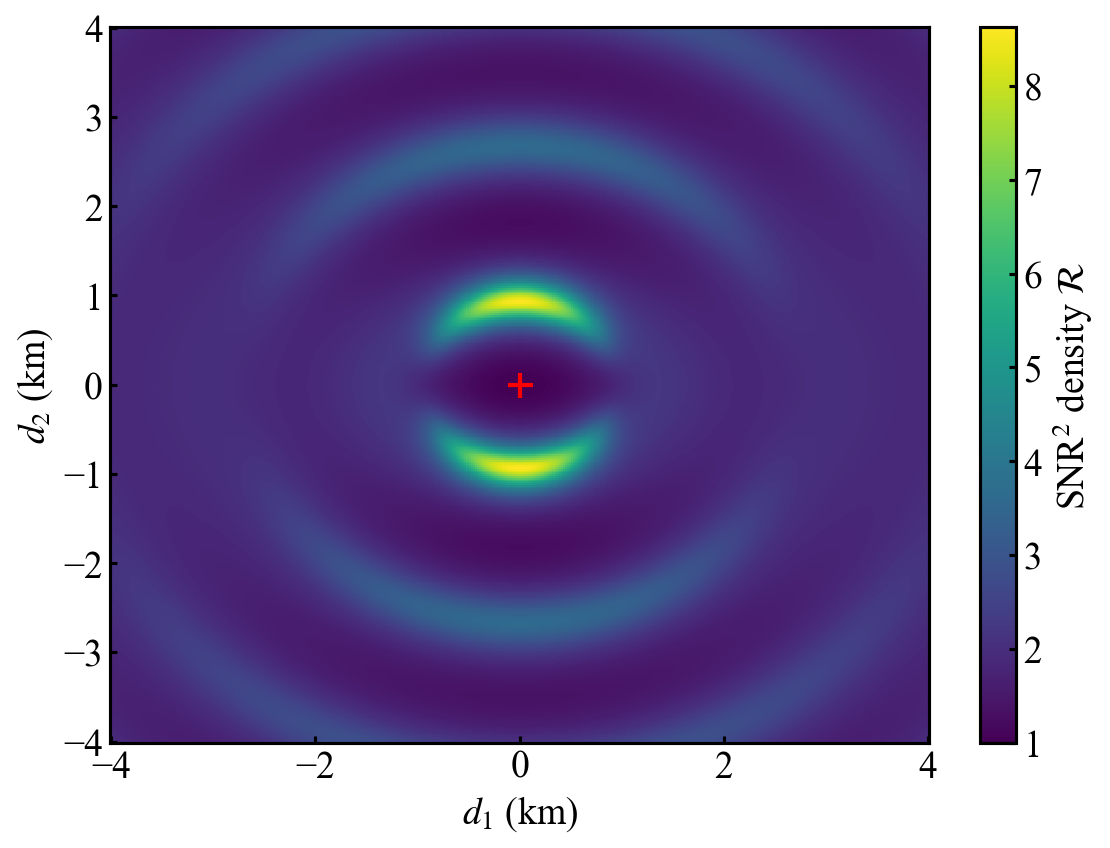}
    \end{subfigure}
    
    \caption{SNR$^2$ density for parallel propagating GW, $(e,\lambda)=(\pi/2,\pi/2)$. $f=0.3$ Hz and $c_R = 500$ m/s. Upper panel: $+$ polarization results, normalized by (2,2) km value. Lower panel: $\times$ polarization results, normalized by (0,0) km value. Normalization locations are marked with red cross. }
    \label{fig:paralGW}
    
\end{figure}

\begin{figure}[htbp] 
    \centering
    
    \begin{subfigure}{}
        \includegraphics[width=\linewidth]{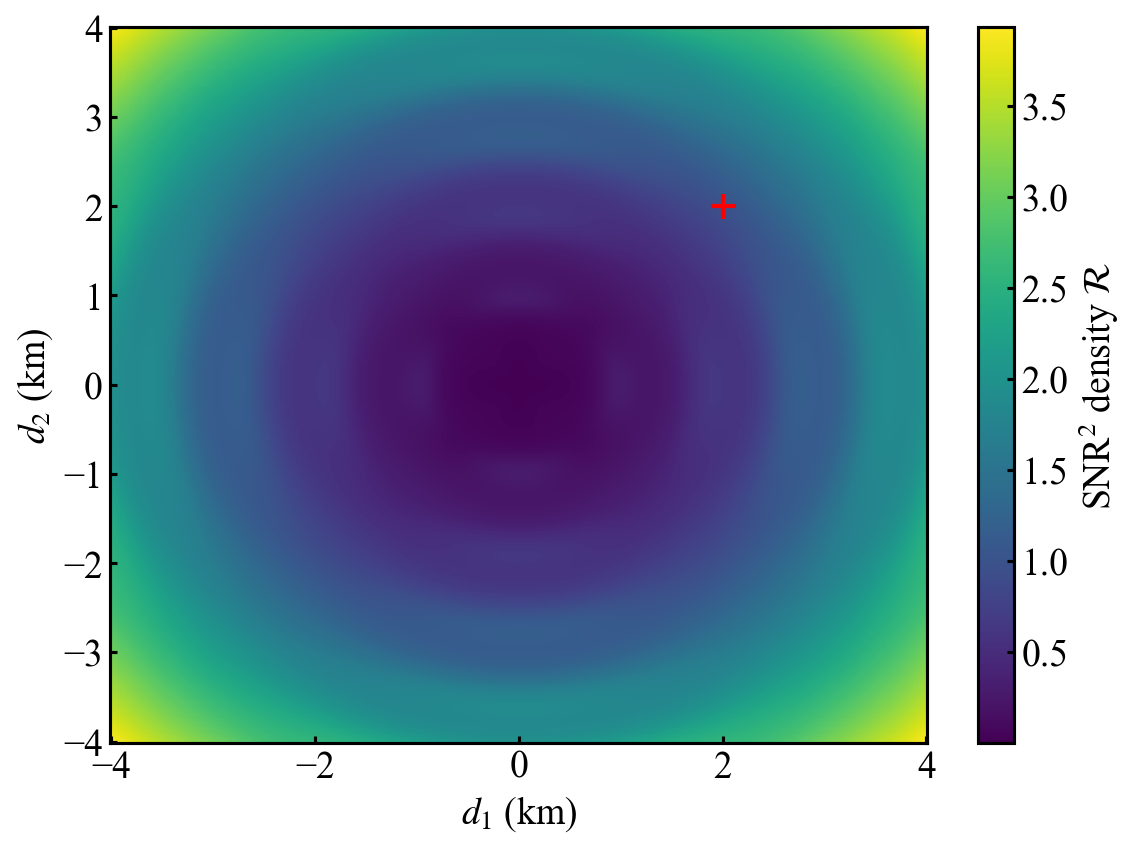}
    \end{subfigure}
    \begin{subfigure}{}
        \includegraphics[width=\linewidth]{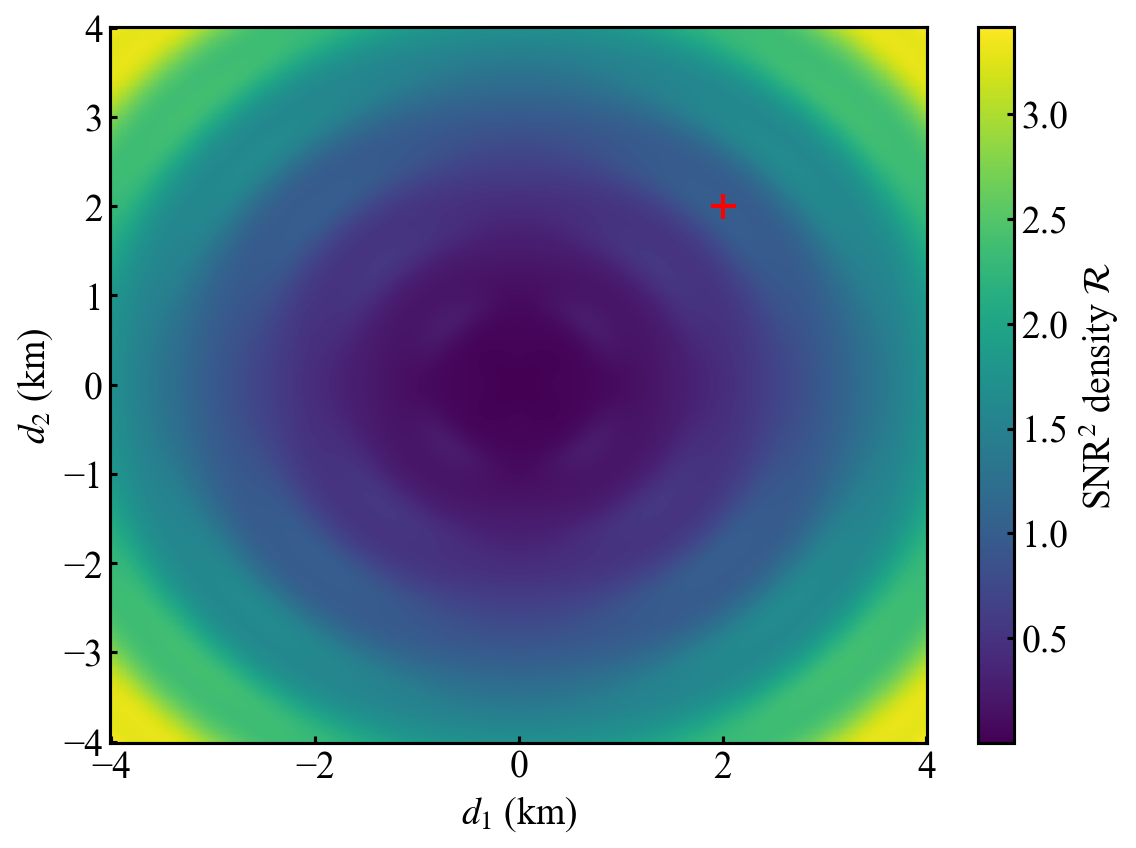}
    \end{subfigure}
    
    \caption{SNR$^2$ density for vertically propagating GW, $(e,\lambda)=(\pi/2,0)$. $f=0.3$ Hz and $c_R = 500$ m/s. Upper panel: $+$ polarization results, normalized by (2,2) km value. Lower panel: $\times$ polarization results, normalized by (2,2) km value. Normalization locations are marked with red cross.}
    \label{fig:vertGW}
    
\end{figure}

In this Appendix, we plot the SNR$^2$ density of a two-station array (same as in Fig.~\ref{fig:fig1}) located at $(\theta,\phi)=(\pi/2,0)$, but for single wave vector direction (specified by $e$ and $\lambda$ in standard spherical coordinates) and polarization state each time. Figure~\ref{fig:paralGW} shows the results for parallel propagating GW, i.e., $(e,\lambda)=(\pi/2,\pi/2)$, in which case GW propagates parallel to the plane formed by sensing directions. Figure~\ref{fig:vertGW} shows the results for vertically propagating GW, i.e., $(e,\lambda)=(\pi/2,0)$ , in which case GW propagates perpendicular to the plane formed by sensing directions.

We note here that for vertically propagating GW, the results for $+$ and $\times$ polarization can be transformed into each other through a rotation of $\pm \pi/4$. However, for parallel propagating GW, two results are completely different.

\section{Test for approximation formula}
\label{app:approxtest}

\begin{figure}[htbp]
\centering
\includegraphics[width=0.95\linewidth]{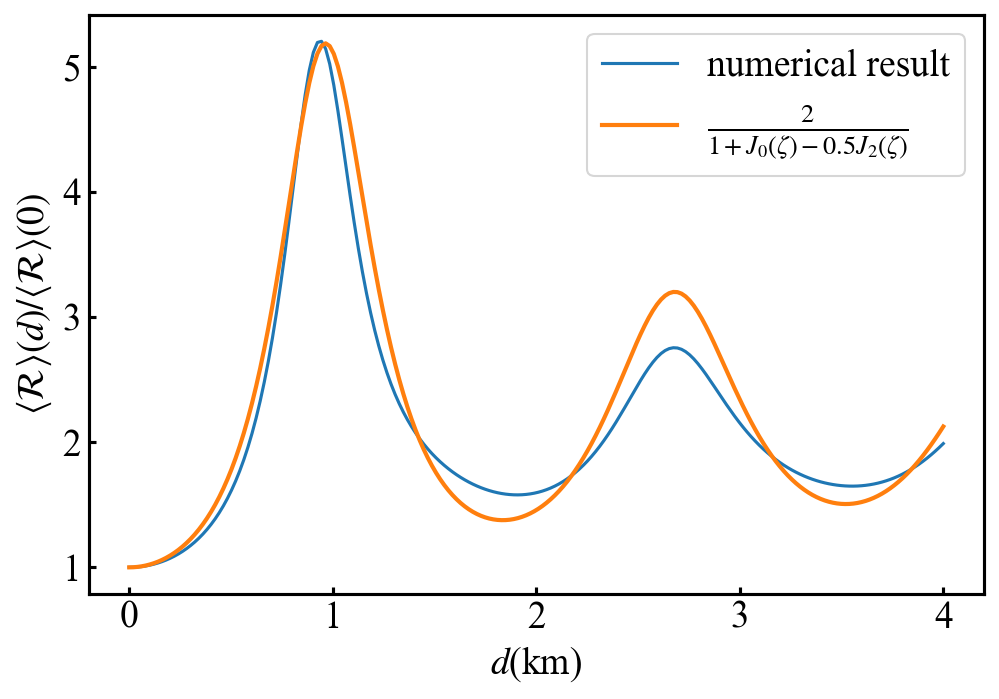}
\caption{Same quantity as in Fig.~\ref{fig:fig2}, compared with analytical approximation formula. The distance is extended to 4~km, while all other parameters remain unchanged. }
\label{fig:Approx}
\end{figure}

In this Appendix, we evaluate the accuracy of the phenomenological approximation given by Eq.~(\ref{eq:averaged_mitigation}). Fig.~\ref{fig:Approx} compares the exact numerical calculation of the normalized sky-averaged $\text{SNR}^2$ density with the analytical formula. Since the numerical results in Fig.~\ref{fig:fig2} demonstrate that the enhancement is independent of the relative orientation between the two stations, the comparison is plotted as a function of a single separation distance $d$.

As shown in Fig.~\ref{fig:Approx}, the analytical approximation closely tracks the numerical results. Across the evaluated distance range (up to 4~km), the maximum relative error between the analytical formula and the exact numerical curve remains below 25\%. Furthermore, the approximation error is nearly negligible in the vicinity of the optimal mitigation peak, confirming its validity for estimating the array's maximum noise-mitigation capability.

\section{Test for a four-station array}
\label{app:four-station}

\begin{figure}[htbp]
\centering
\includegraphics[width=0.95\linewidth]{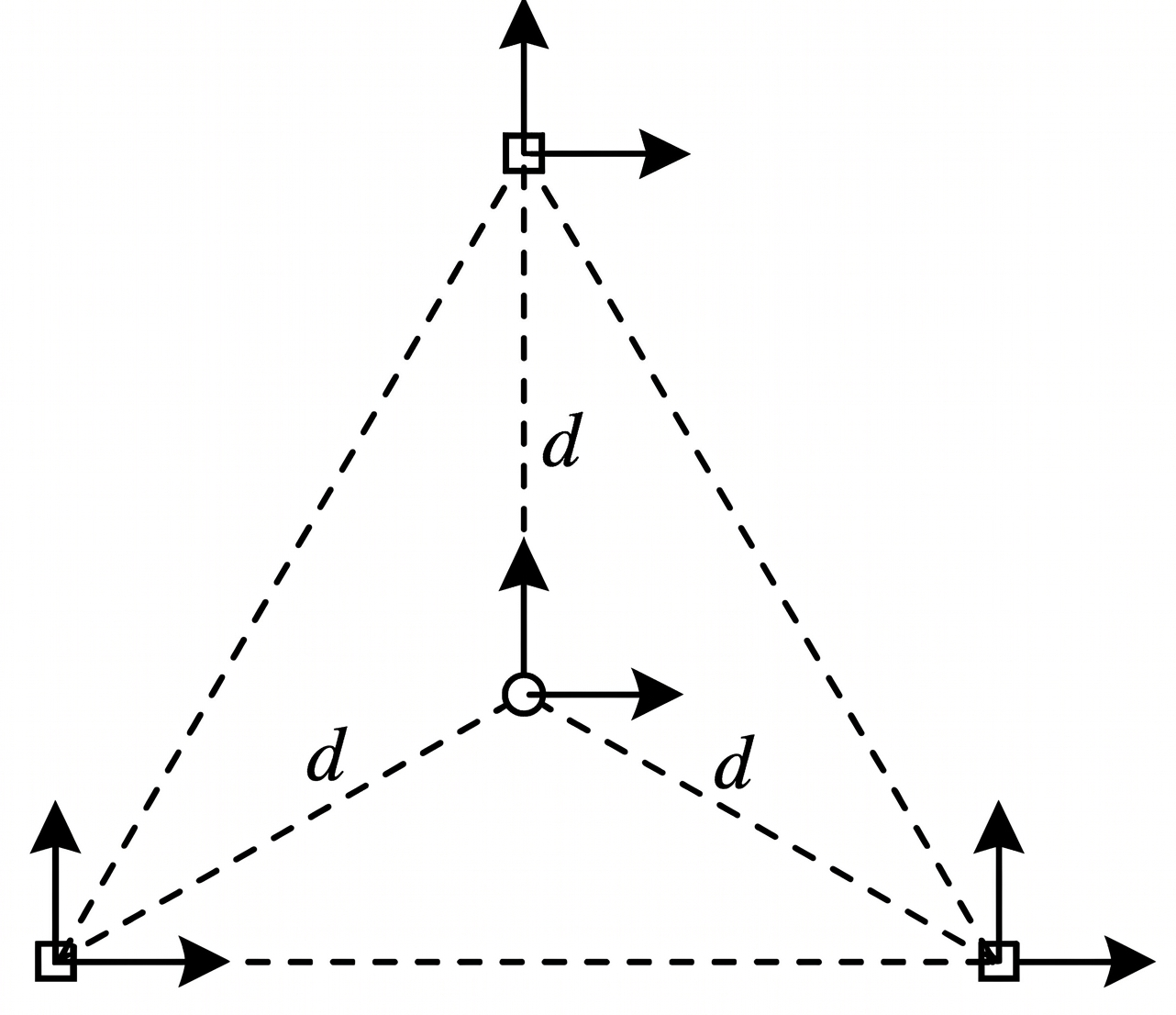}
\caption{Schematic of the typical ``four-station array'' configuration. Three stations are located at the vertices of an equilateral triangle, and one at the centroid. }
\label{fig:4stat}
\end{figure}

To further illustrate the potential mitigation capabilities of a larger network, we evaluate a representative four-station array configuration. As shown in Fig.~\ref{fig:4stat}, three stations are positioned at the vertices of an equilateral triangle, with the fourth station located at the geometric centroid. Each of them has two mutually perpendicular sensors. We maintain the assumption that all inter-station distances are much smaller than the lunar radius.

\begin{figure}[htbp]
\centering
\includegraphics[width=0.95\linewidth]{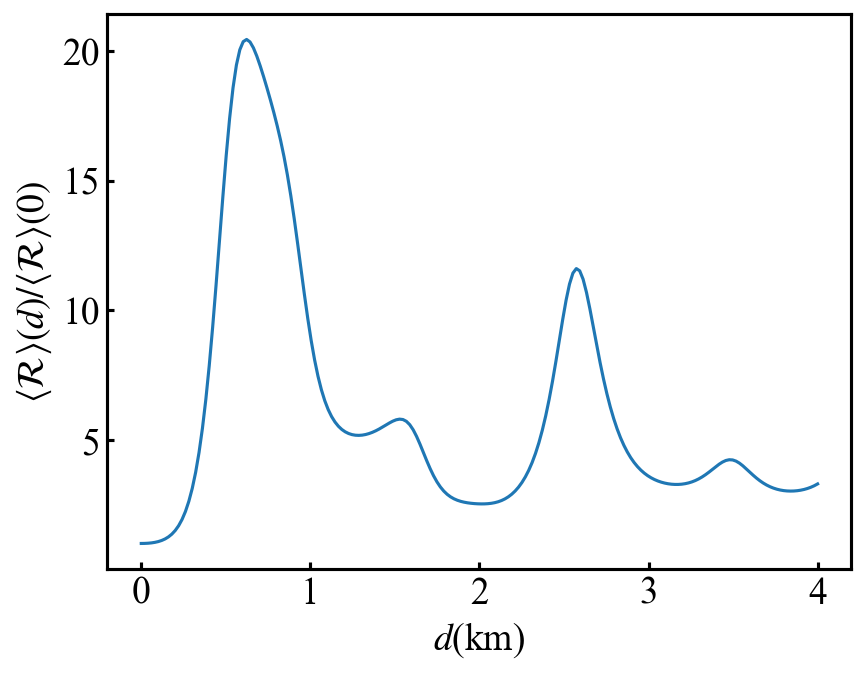}
\caption{Sky-averaged SNR$^2$ density for the four-station array, normalized by zero-separation ($d=0$) value. $f=0.3~\text{Hz}$ and $c_R=500~\text{m/s}$. The horizontal axis indicates the distance from the centroid to the vertices. }
\label{fig:4stat-mit}
\end{figure}

Fig.~\ref{fig:4stat-mit} presents the normalized sky-averaged SNR$^2$ density for this four-station configuration, evaluated at the same physical parameters used in Fig.~\ref{fig:fig2} ($f = 0.3$ Hz and $c_R = 500$ m/s). The horizontal axis denotes the distance from the centroid to the vertices (which proportionally scales the entire array), and the vertical axis represents the normalized sky-averaged SNR$^2$ density in the seismic-noise-
dominated regime. The results reveal that the maximum SNR$^2$ density reaches approximately 20. This corresponds to a significant enhancement in the equivalent noise ASD mitigation, achieving a factor of roughly $1/\mathcal{M}_{\text{eff}} \sim \sqrt{20} \approx 4.5$, which noticeably outperforms the two-station scenario. Furthermore, in the large-separation limit, the normalized SNR$^2$ density asymptotically approaches 4, consistently matching the expected behavior of four uncorrelated stations.

Although this preliminary calculation does not constitute a formal, comprehensive array design study for LGWA, which would require incorporating additional complexities such as body-wave contributions, non-isotropic background components, and detailed topographical models, it clearly demonstrates the enhanced mitigation capacity of multi-station networks. This highlights the critical importance of further array geometry optimization to identify the most effective parameter space for seismic noise mitigation.

\clearpage


\end{document}